\documentclass{article}
\usepackage{spconf,amsmath,graphicx,hyperref,booktabs}
\usepackage{balance, multirow}
\usepackage{booktabs}
\usepackage{makecell}
\usepackage{graphicx}
\usepackage{threeparttable}

\title{The Second MLC-SLM Challenge: Multilingual Conversational Speech Diarization, Recognition, and Understanding}
\name{
\parbox{\linewidth}{\centering
Bingshen Mu$^1$, Mingchen Shao$^1$, Zhennan Lin$^1$, Liumeng Xue$^2$, Hexin Liu$^3$,\\Lei Xie$^{1*}$, Eng Siong Chng$^3$, Longshuai Xiao$^4$, Qiangze Feng$^5$, Daliang Wang$^5$}\thanks{$^*$Corresponding author.}}
\address{$^1$Audio, Speech and Language Processing Group (ASLP@NPU), School of Computer Science, \\Northwestern Polytechnical University, Xi'an, China\\
$^2$School of Intelligence Science and Technology, Nanjing University\\
$^3$College of Computing and Data Science, Nanyang Technological University, Singapore\\
$^4$Huawei Technologies, China\\$^5$Nexdata Technology Inc., USA
}

\begin{document}
\ninept
\maketitle
\begin{abstract}
\vspace{-4.5pt}
This paper summarizes the Interspeech2026 second Multilingual Conversational Speech Language Model (MLC-SLM) Challenge, which aims to advance the development of effective multilingual conversational speech language models. We describe the two challenge tasks: multilingual conversational speech diarization and recognition, and multilingual conversational speech understanding, together with the released real-world conversational speech dataset, evaluation protocols, and baseline systems. The challenge attracted 91 teams worldwide, with 704 valid leaderboard results and 14 technical reports across the two tasks. Based on the participating systems, we summarize representative approaches and distill practical insights into multilingual conversational speech recognition and understanding to support future research in the community.
\end{abstract}
\vspace{-9pt}
\section{Introduction}
\vspace{-9pt}
\label{sec:intro}
Large language models (LLMs) have demonstrated strong capabilities in language understanding, generation, and reasoning~\cite{grattafiori2024llama, team2024gemma, yang2025qwen3}. Recent studies have extended these capabilities to speech, driving the rapid development of speech LLMs (SLLMs). By integrating speech encoders with LLMs, SLLMs have achieved promising results in automatic speech recognition (ASR)~\cite{mu2024mmger, mu2025hdmole, mu2025mixture, shi2026qwen3}, speech synthesis~\cite{anastassiou2024seed, wang2025spark, hu2026qwen3}, speech enhancement~\cite{wang2024selm, kang2025llase}, and spoken dialogue systems~\cite{xu2025qwen2, wang2024freeze, defossez2024moshi}. However, building robust multilingual conversational SLLMs requires real-world speech corpora that reflect the complexity of natural communication, including speaker turns, interruptions, overlapped speech, diverse accents, acoustic variations, and long-range contextual dependencies. Such data remain scarce, particularly for multilingual conversational scenarios.

The Interspeech2025 first Multilingual Conversational Speech Language Model (MLC-SLM) Challenge~\cite{mu2026summary} promotes research on multilingual conversational speech modeling by releasing a large-scale real-world conversational speech corpus and defining tasks for multilingual speech recognition and speaker-attributed transcription. Its results demonstrate the effectiveness of SLLMs for multilingual conversational speech recognition. Nevertheless, accurately determining who speaks when remains challenging in complex multilingual conversations. Moreover, transcription accuracy alone is insufficient for evaluating whether a model can understand the acoustic events, speaker relationships, and semantic content of an entire conversation. These limitations motivate the second MLC-SLM Challenge, which expands both linguistic coverage and task scope from conversational speech transcription to speaker-aware recognition and holistic spoken language understanding.

The Interspeech2026 second MLC-SLM Challenge\footnote{https://www.nexdata.ai/competition/mlc-slm} releases around 2,100 hours of natural two-speaker conversations covering 14 languages and 21 language and accent variants. It introduces 3 additional languages, including Tagalog, Urdu, and Turkish, and extends the coverage of regional varieties, including Canadian French, Mexican Spanish, and Brazilian Portuguese. The challenge consists of two tasks. Task 1 requires systems to jointly perform multilingual conversational speech diarization and recognition from unsegmented audio without oracle speaker labels or utterance boundaries. Task 2 evaluates multilingual conversational speech understanding through multiple-choice questions concerning acoustic, semantic, and joint acoustic-semantic information in complete conversations. The challenge attracts 91 teams worldwide, resulting in 704 valid leaderboard results and 14 technical reports across both tasks. In this paper, we describe the released dataset, task settings, evaluation protocols, and baseline systems, and summarize representative methods developed by the participating teams. We further distill practical insights for building effective multilingual conversational SLLMs and discuss the remaining problems revealed by the challenge.

\vspace{-9pt}
\section{Task Settings and Rules}
\vspace{-9pt}
\label{sec:setting}
\textbf{Task 1: Multilingual Conversational Speech Diarization and Recognition.} This task aims to jointly determine who speaks when and transcribe what each speaker says in multilingual conversations. The input consists of complete two-speaker conversational recordings. Oracle segmentation and speaker labels are provided for the training (Train) set, whereas the evaluation recordings are released without utterance boundaries, timestamps, speaker labels, or other oracle information. Participants must generate speaker-attributed transcriptions containing speaker identities, time boundaries, and recognized text. Both cascaded and end-to-end systems are allowed.
The systems are evaluated using diarization error rate (DER) and time-constrained minimum-permutation word error rate or character error rate (tcpWER/tcpCER). We use tcpCER for Japanese, Korean, and Thai, and tcpWER for the other languages. For each recording, the hypotheses and reference transcriptions assigned to the same speaker are concatenated after speaker permutation, and the corresponding word or character error rate is calculated. The final ranking is determined by time-constrained minimum-permutation mixed error rate (tcpMER), with lower values indicating better performance.

\textbf{Multilingual Conversational Speech Understanding.} This task evaluates whether a system can understand the acoustic and semantic information conveyed in a complete multilingual conversation. Each evaluation instance consists of a conversational recording and a multiple-choice question with two to four candidate answers, only one of which is correct. The questions cover three categories: acoustic information, semantic content, and joint acoustic-semantic information. As in Task 1, no oracle segmentation, timestamps, or speaker labels are provided during evaluation.
Task 2 uses the same conversational speech corpus as Task 1. However, no dedicated multiple-choice Train set is officially released. Participants may construct training questions and answers based on the released speech data and the examples provided in the development (Dev) set. Both pipeline-based and end-to-end systems are permitted. System performance is measured by answer accuracy; higher values indicate better performance. The evaluation is conducted in two phases, and only the Phase 2 result is used to determine the final ranking.

\textbf{Rules.} Systems submitted to either task must be developed based on LLMs, SLLMs, or multimodal LLMs. Publicly and freely accessible external datasets, speech foundation models, and LLMs are permitted. Data augmentation methods, including additive noise, reverberation, speed perturbation, and pitch modification, are also allowed.
The evaluation (Eval) set must not be used for training, fine-tuning, parameter selection, or any other operation that may cause leakage. Multi-system output fusion, such as combining the hypotheses of multiple ASR systems using ROVER~\cite{fiscus1997post}, is prohibited. Nevertheless, a single pipeline composed of multiple functional components, such as voice activity detection (VAD), diarization, and ASR, is allowed. Commercial APIs may only be used to construct multiple-choice training data for Task 2 and must not be used to generate predictions on the Eval set. 

\vspace{-9pt}
\section{Released Dataset}
\vspace{-9pt}
\label{sec:dataset}
The second MLC-SLM Challenge releases approximately 2,100 hours of multilingual conversational speech. The Train set contains 7,080 natural two-speaker conversations recorded at 16 kHz, mainly in quiet indoor environments using mobile devices. The speakers cover diverse age groups and genders, and each conversation focuses on a randomly assigned topic.
The corpus covers 14 languages and 21 language and accent variants. Compared with the first challenge, Tagalog, Urdu, and Turkish are newly included, together with Canadian French, Brazilian Portuguese, and Mexican Spanish. The English data cover American, British, Filipino, Australian, and Indian English. The Train and Dev sets provide transcriptions, oracle segmentation, timestamps, and speaker labels. Task 2 uses the same conversational recordings, but no dedicated multiple-choice Train set is provided.
The Dev set contains 150 long-form conversations and 4,500 multiple-choice questions for Task 2. During evaluation, the recordings are released without oracle segmentation, timestamps, speaker labels, or reference transcriptions. Task 2 is evaluated in two phases, and its final Phase 2 set contains 9,470 questions. The questions assess acoustic, semantic, and joint acoustic-semantic understanding and are manually reviewed to ensure clarity and a unique correct answer.

\vspace{-9pt}
\section{Baselines}
\vspace{-9pt}
\label{sec:baselines}
The baseline systems are developed separately for the two tasks. For Task 1, we fine-tune Microsoft’s open-source VibeVoice-ASR~\cite{peng2026vibevoice} on the challenge Train set\footnote{https://github.com/alanshaoTT/MLC-SLM-2nd-Task1-Baseline}. VibeVoice-ASR uses a speech encoder to encode the input audio and a Qwen2.5-7B LLM to generate speaker-attributed transcriptions containing timestamps, speaker labels, and recognized text. The speech encoder is frozen, while Low-Rank Adaptation (LoRA)~\cite{hu2022lora} is applied to the attention and feed-forward layers of the LLM. During evaluation, tcpWER/tcpCER is computed using the Meeteval toolkit\footnote{https://github.com/fgnt/meeteval} with a collar of 5 seconds. Japanese, Korean, and Thai are evaluated using tcpCER, while the other languages are evaluated using tcpWER. The baseline achieves an average tcpMER of 79.15\% on the Dev set.

For Task 2, the baseline is built on Qwen2.5-Omni-7B~\cite{xu2025qwen2}, an omni-modal LLM capable of directly processing speech and text\footnote{https://github.com/DontPushMeee/MLC-SLM-2nd-Task2-Baseline}. Gemini2.5-Pro is used to generate questions covering acoustic, semantic, and joint acoustic-semantic information from the released training conversations. The model is fine-tuned with LoRA using the ms-swift framework with a Megatron-LM backend. During inference, the complete conversation, question, and candidate answers are provided to the model, which generates the selected option directly. A retry mechanism and sampling strategy are used to improve answer extraction. On a development subset containing 300 multiple-choice questions, the baseline obtained an overall accuracy of 35.33\%. Its accuracy is 47.75\% for acoustic questions, 33.67\% for joint acoustic-semantic questions, and 21.98\% for semantic questions, indicating that conversation-level semantic reasoning remained particularly challenging.
\begin{table*}[t]
\centering
\caption{Results and main techniques of the top-ranked Task 1 systems with valid technical reports. Dev and Eval results are average tcpMER values, expressed as percentages. A dash indicates that the corresponding information is not reported.}
\label{tab:task1_methods}
\begin{threeparttable}
\resizebox{0.965\textwidth}{!}{
\begin{tabular}{c l c l l l l c}
\toprule
\textbf{Rank} &
\textbf{Team} &
\textbf{Architecture} &
\textbf{Diarization} &
\textbf{ASR/SLLM} &
\textbf{Training and adaptation} &
\textbf{External data} &
\textbf{Eval} \\
\midrule

1 &
MOSS Transcribe Diarize &
End-to-end &
\makecell[l]{Joint speaker-token\\generation} &
\makecell[l]{Whisper-large-v3\\+ Qwen3-8B} &
\makecell[l]{Three-stage training; speaker-attributed\\SFT; domain adaptation} &
\makecell[l]{Multilingual ASR, meeting,\\and synthetic data} &
\textbf{14.93} \\

2 &
trasr~\cite{trasr} &
Cascaded &
\makecell[l]{DiariZen; VoxBlink2;\\VBx/PLDA} &
\makecell[l]{Qwen3-Omni-30B\\+ Qwen3-ASR} &
\makecell[l]{Diarization adaptation; two-stage\\LoRA; checkpoint averaging} &
\makecell[l]{Public diarization\\corpora} &
\textbf{15.41} \\

3 &
Cake by VPBank &
End-to-end &
\makecell[l]{Joint prediction by\\VibeVoice-ASR} &
VibeVoice-ASR &
\makecell[l]{Label recovery; merge-then-fresh\\LoRA; silence/noise augmentation} &
\makecell[l]{External transcription model;\\noise corpora} &
\textbf{15.84} \\

4 &
HINTT &
Cascaded &
\makecell[l]{DiariZen;\\SimAMResNet100} &
\makecell[l]{Qwen3-ASR-1.7B\\+ Qwen3.6-27B} &
\makecell[l]{Full fine-tuning; SpecAugment;\\N-best error correction} &
Official data only &
\textbf{16.27} \\

5 &
roysun2006~\cite{roysun2006} &
End-to-end &
\makecell[l]{Joint prediction by\\VibeVoice-ASR} &
VibeVoice-ASR-7B &
\makecell[l]{LoRA; leading-silence\\cropping; EMA} &
-- &
\textbf{16.73} \\

6 &
SQZ~\cite{sqz} &
Cascaded &
\makecell[l]{FSMN-VAD; CAM++;\\spectral clustering} &
Qwen3-ASR-1.7B &
\makecell[l]{Full SFT; synthetic-speech\\LoRA; GRPO} &
\makecell[l]{Common Voice;\\synthetic speech} &
\textbf{17.97} \\

7 &
xdynamics~\cite{xdynamics} &
Cascaded &
\makecell[l]{FSMN-VAD; ReDimNet2;\\spectral clustering} &
Qwen3-ASR-1.7B &
LoRA adaptation &
-- &
\textbf{18.44} \\

8 &
SMIIP Lab &
\makecell[c]{Diarization-\\conditioned} &
w2v-S2SND &
\makecell[l]{Whisper-large-v3-turbo\\+ Qwen3-4B} &
\makecell[l]{Simulated pretraining; projector\\training; LoRA} &
$\sim$4,800 h &
\textbf{19.97} \\

9 &
fangshuming~\cite{fangshuming} &
Cascaded &
\makecell[l]{FSMN-VAD; DiariZen;\\CAM++} &
omniASR LLM 7B v2 &
\makecell[l]{LoRA; speaker-change tokens;\\language-constrained decoding} &
-- &
\textbf{50.23} \\

\bottomrule
\end{tabular}
}
\end{threeparttable}
\vspace{-9pt}
\end{table*}

\vspace{-9pt}
\section{Methods of Task 1}
\vspace{-9pt}
\label{sec:methods1}
Nine top-ranked teams with valid technical reports are summarized in Table~\ref{tab:task1_methods}. The submitted systems can be broadly divided into three categories: end-to-end systems that directly generate speaker-attributed transcriptions, cascaded systems that connect diarization and ASR modules, and diarization-conditioned systems that provide explicit speaker and temporal information to the ASR model. The best evaluation results are achieved by both end-to-end and cascaded systems, indicating that performance depends not only on the overall architecture but also on speaker modeling, temporal alignment, training-data quality, and long-form inference.
\vspace{-9pt}
\subsection{System Architecture}
\vspace{-4.5pt}
The end-to-end systems represented by MOSS Transcribe Diarize, Cake by VPBank, and roysun2006 directly generate timestamps, speaker labels, and transcriptions from complete recordings. MOSS Transcribe Diarize combines a Whisper-large-v3 encoder~\cite{radford2023robust}, an acoustic adaptor, and Qwen3-8B, and represents speaker identities and temporal positions as tokens in a unified output sequence. The two VibeVoice-ASR systems use a similar unified formulation without an external diarization module. This design avoids hard error propagation between diarization and ASR, but requires the generative model to maintain stable speaker identities and timestamps throughout a long recording.
Cascaded systems remain highly competitive. Trasr combines a fine-tuned diarization system with Qwen3-Omni and CTC-based timestamp estimation, achieving the second-best result. HINTT applies diarization, segment-level Qwen3-ASR recognition, and LLM-based N-best correction. SQZ, xdynamics, and fangshuming employ modular pipelines consisting of VAD, speaker embedding extraction, clustering, segmentation, and ASR. SMIIP adopts an intermediate design in which a diarization model predicts speaker turns and boundaries that are converted into discrete speaker and time tokens for the downstream SLLMs.
\vspace{-9pt}
\subsection{Speaker Diarization and Attribution}
\vspace{-4.5pt}
Speaker attribution is handled either implicitly within an end-to-end model or explicitly through a diarization pipeline. The end-to-end systems generate recording-local speaker tokens jointly with timestamps and text. To prevent dependence on fixed speaker identities, MOSS Transcribe Diarize randomly remaps speaker tokens during training.
Most cascaded systems use a local-to-global diarization procedure. Trasr and HINTT fine-tune DiariZen~\cite{han2025leveraging} on the challenge data and combine it with speaker embeddings and clustering. Trasr further assigns ASR tokens to speaker segments according to temporal overlap, using a nearest-segment rule when no overlap is found. SQZ, xdynamics, and fangshuming use VAD followed by short-window speaker embeddings and spectral clustering, with the number of speakers fixed to two. The reported ablations show that in-domain diarization adaptation and speaker-embedding refinement substantially reduce the final recognition error, making diarization one of the most influential components in cascaded systems.
\vspace{-9pt}
\subsection{Training and Model Adaptation}
\vspace{-4.5pt}
The submitted systems employ full fine-tuning, LoRA, or combinations of multiple training stages. MOSS Transcribe Diarize first activates multilingual ASR capabilities using large-scale speech data, then performs speaker-attributed supervised fine-tuning, and finally adapts the system to the official challenge data. Trasr uses two-stage LoRA adaptation, with a higher-rank adapter and targets speed perturbation in the second stage for weaker languages. Cake by VPBank merges an existing LoRA adapter into the backbone and trains a fresh adapter to reduce accumulated parameter drift.
HINTT fully fine-tunes both its diarization and ASR models and uses an external LLM to select or minimally correct N-best hypotheses. Roysun2006 combines LoRA with exponential moving average parameters. SQZ first performs supervised fine-tuning, then adapts the model using synthetic speech and finally applies GRPO~\cite{grpo} with recognition- and hallucination-aware rewards. Its ablation results indicate that supervised fine-tuning contributes most of the improvement, while synthetic-data adaptation and reinforcement learning provide smaller additional gains.

\vspace{-9pt}
\subsection{Data Augmentation and Data Quality}
\vspace{-4.5pt}
The systems use both conventional speech augmentation and task-specific data construction. MOSS Transcribe Diarize synthesizes multi-speaker conversations with realistic turns, pauses, overlaps, and timestamps. Trasr uses noise and speed perturbation, while HINTT applies SpecAugment. SQZ generates multilingual synthetic speech using several text-to-speech strategies and incorporates it through LoRA adaptation.
Several teams find that data quality is as important as data quantity. Cake by VPBank detects speech regions missing from the official annotations, recovers reliable transcriptions, and silences uncertain regions. It also introduces short-silence removal and additive noise to improve robustness. HINTT and trasr similarly mask audible regions outside the annotated intervals. These approaches reduce the mismatch between audio and supervision and help prevent deletion errors or spurious transcriptions.
\vspace{-18pt}
\subsection{Long-form Inference and Post-processing}
\vspace{-4.5pt}
Long conversational recordings introduce timestamp drift, repetition, hallucination, and cross-chunk speaker inconsistency. MOSS Transcribe Diarize inserts periodic time markers into the audio representation and uses a long-context LLM to generate the complete speaker-attributed transcript. When repetitive tails are detected, the system falls back to greedy decoding. Cake by VPBank processes most recordings in a single pass and uses overlapping chunks for longer recordings, followed by voting to align speaker identities across chunks.
Cascaded systems rely on more explicit temporal processing. Trasr generates word- or character-level timestamps with a CTC head and aligns recognized tokens with diarization segments. HINTT retains beam-search N-best hypotheses and applies conservative LLM-based error correction. Other systems use repetition detection, language-constrained decoding, text normalization, speaker-turn merging, and duplicate removal. Overall, the results suggest that robust long-form decoding and accurate speaker–time–text alignment are more decisive than the choice between end-to-end and cascaded architectures alone.
\begin{table*}[t]
\centering
\caption{Results and main techniques of the Task 2 systems with valid technical reports. Dev and Eval results are accuracies in percent. A dash indicates that the corresponding result is not reported.}
\label{tab:task2_methods}
\begin{threeparttable}
\resizebox{0.92\textwidth}{!}{
\begin{tabular}{c l l l l l c}
\toprule
\textbf{Rank} &
\textbf{Team} &
\textbf{Evidence and architecture} &
\textbf{Backbone} &
\textbf{Task-specific supervision} &
\textbf{Model adaptation} &
\textbf{Phase 2 Eval} \\
\midrule

1 &
hfchen &
\makecell[l]{Audio-native end-to-end\\reasoning} &
MOSS-Audio-8B-Thinking &
\makecell[l]{Synthetic QA with\\consistency filtering} &
\makecell[l]{SFT + DAPO-style\\reinforcement learning} &
\textbf{93.9} \\

2 &
ClaudeAKE~\cite{le2026sear} &
\makecell[l]{Audio-native reasoning over\\localized event segments} &
Qwen3-Omni-30B-A3B &
\makecell[l]{359,825 verified acoustic\\and semantic QA pairs} &
\makecell[l]{LoRA SFT\\+ GSPO} &
\textbf{90.9} \\

3 &
xdynamics~\cite{xdynamics} &
\makecell[l]{Transcript-assisted dynamic\\evidence routing} &
Qwen3-Omni-30B-A3B &
\makecell[l]{No additional Task 2\\training data} &
\makecell[l]{Training-free routing\\and answering} &
\textbf{90.7} \\

4 &
roysun2006~\cite{roysun2006} &
\makecell[l]{Audio-native end-to-end\\answering} &
Qwen3-Omni-30B-A3B &
\makecell[l]{$\sim$127,000 filtered and\\augmented QA pairs} &
LoRA SFT &
\textbf{86.0} \\

5 &
eloquence~\cite{luque2026eloquence} &
\makecell[l]{Multimodal in-context\\learning} &
Voxtral-24B &
\makecell[l]{Balanced multimodal\\demonstrations} &
Training-free ICL &
\textbf{80.9} \\

\bottomrule
\end{tabular}
}
\end{threeparttable}
\vspace{-9pt}
\end{table*}

\vspace{-9pt}
\section{Methods of Task 2}
\vspace{-9pt}
Table~\ref{tab:task2_methods} summarizes the five Task 2 systems for which technical reports are available. The submitted approaches can be broadly characterized by how they represent conversational evidence, construct task-specific supervision, adapt multimodal LLMs, and control answer generation. The leading systems primarily use audio-native multimodal models, whereas other systems combine speaker-attributed transcripts with question-dependent audio evidence or adopt training-free in-context learning.
\vspace{-9pt}
\subsection{Evidence Representation and System Architecture}
\vspace{-4.5pt}
Hfchen and roysun2006 adopt audio-native end-to-end systems that directly process the complete conversational recording together with the question and candidate answers. MOSS-Audio-8B-Thinking combines an audio encoder with Qwen3-8B and uses explicit temporal markers and cross-layer audio injection to support long-form speech reasoning. The system generates an intermediate reasoning trace before producing the final answer. Roysun2006 uses Qwen3-Omni-30B-A3B to directly answer questions from complete audio without an intermediate ASR transcript.
ClaudeAKE also uses an audio-native multimodal model for prediction, but localizes relevant events during training-data construction using timestamped ASR and language-model-based event detection. Xdynamics instead treats speaker-attributed transcription as the primary evidence and dynamically selects among full transcripts, localizes audio–text evidence, speaker-linked evidence, and globally samples audio according to the question type. Its development experiments show that adding localized audio to the transcript backbone improves accuracy from 92.69\% to 94.22\%, while the final hybrid router achieves 94.84\%.
Eloquence explores LoRA fine-tuning, multimodal in-context learning, and transcript-assisted retrieval. Its best submission uses a frozen Voxtral-24B model with balanced multimodal demonstrations. This approach increases Phase 2 accuracy from 74.3\% without demonstrations to 81.0\% with six multimodal examples, showing that carefully selected demonstrations can reduce answer-position bias without parameter updates.
\vspace{-9pt}
\subsection{Task-specific Supervision Construction}
\vspace{-4.5pt}
Because the challenge does not provide a dedicated multiple-choice Train set, most teams construct supervision from the released conversational recordings. Hfchen generates questions from transcripts and speaker labels, covering spoken content, speaker identity, turn order, temporal relations, and conversation-level intent. The generated examples are filtered according to answerability, answer uniqueness, label validity, and consistency with the source conversation.
ClaudeAKE separately generates semantic and acoustic questions. Semantic questions are created from localized events using a text LLM, whereas acoustic questions are generated directly from waveform segments using a multimodal model. After multi-stage verification, the resulting dataset contains 359,825 examples. A text-only probe further divides them into 197,231 weak examples that can be answered primarily from textual cues and 162,594 strong examples that require acoustic evidence.
Roysun2006 initially generates approximately 210,000 questions and applies a silent-audio counterfactual test to remove examples that do not require speech input. Approximately 67,000 examples remain after filtering, and around 60,000 additional examples are generated to better match the evaluation distribution. Its cumulative results improve from 81.0\% before audio-dependency filtering to 83.0\% after filtering and to 85.0\% after distribution-matched augmentation. These results indicate that the relevance and acoustic dependence of synthetic supervision are more important than its raw size.
\vspace{-9pt}
\subsection{Model Adaptation and Learning Objectives}
\vspace{-4.5pt}
The two highest-ranked systems combine supervised fine-tuning with task-specific reinforcement learning. Hfchen first performs supervised adaptation on challenge-style questions and then applies DAPO-style reinforcement learning using exact answer correctness as the main reward. ClaudeAKE uses the weak examples for LoRA-based supervised fine-tuning and the strong audio-dependent examples for GSPO~\cite{gspo}. Its routing-based training strategy achieves 90.92\%, compared with 88.24\% for a control system trained without separating weak and strong examples.
Roysun2006 uses two-stage LoRA supervised fine-tuning while allowing the visual encoder and modality aligner to be updated. Xdynamics does not fine-tune its Task 2 router or answering model and instead relies on deterministic cues and prompt-based evidence allocation. Eloquence shows that training-free multimodal in-context learning can remain competitive when the demonstrations are balanced across answer labels, although its smaller LoRA-fine-tuned model exhibits a larger development-to-evaluation performance gap.
\vspace{-9pt}
\subsection{Inference and Answer Control}
\vspace{-4.5pt}
The systems differ in whether they process complete recordings or question-dependent evidence. Hfchen and roysun2006 use complete audio, while ClaudeAKE operates on localized event segments and Xdynamics dynamically selects the evidence scope and modality. Eloquence uses fixed short audio crops as multimodal demonstrations together with the evaluation recording and question.
Strict answer formatting is widely adopted to improve evaluation reliability. ClaudeAKE and roysun2006 require the selected option to appear within an explicit answer tag, while hfchen applies deterministic parsing and verifies that the generated answer belonged to the candidate set. Eloquence constrains the model to output a single option letter and balances the labels represented in its demonstrations. The reported results suggest that, beyond acoustic and semantic modeling, audio-dependent supervision, question-conditioned evidence selection, and reliable answer extraction are central to multilingual conversational speech understanding.

\vspace{-9pt}
\section{Conclusion}
\vspace{-9pt}
This paper summarizes the second MLC-SLM Challenge, including its dataset, task settings, baseline systems, and representative submissions. The challenge results demonstrate the effectiveness of both end-to-end and cascaded systems for multilingual conversational speech diarization and recognition, while audio-dependent supervision and task-specific optimization are important for conversational speech understanding.
We hope that the challenge resources and findings will facilitate further research on robust multilingual conversational SLLMs.
Cross-system comparisons are observational because the submitted systems
differ in model scale, external data, synthetic supervision, and computational resources. Future editions should consider controlled-data tracks and report results by language, overlap condition, and question category to enable more fine-grained and reproducible comparisons.
\clearpage
\balance
\bibliographystyle{IEEEbib}
\bibliography{refs}

\end{document}